\documentclass[aps,prl,twocolumn,showpacs,groupedaddress]{revtex4}

\usepackage{amsmath}
\usepackage{graphicx}
\usepackage{dcolumn}
\usepackage{bm}

\begin{document}


\title{Quantum Monte Carlo simulation of overpressurized liquid $^{\bf
4}$He}

\author{L. Vranje\v{s}}
\affiliation{Faculty of Natural Sciences, University of Split, 21000 Split,
Croatia}
\author{J. Boronat}
\author{J. Casulleras}
\author{C. Cazorla}
\affiliation{Departament de F\'\i sica i Enginyeria Nuclear, Campus Nord
B4-B5, Universitat Polit\`ecnica de Catalunya, E-08034 Barcelona, Spain}

\date{\today}
\begin{abstract}
 A diffusion Monte Carlo simulation of superfluid $^4$He at zero
 temperature and pressures up to 275 bar is presented. Increasing the
 pressure beyond freezing ($\sim$ 25 bar), the liquid enters the 
 overpressurized phase in a
 metastable state. In this regime, we report results of the equation of
 state and the pressure dependence of the static structure factor, the
 condensate fraction, and the excited-state energy corresponding to the
 roton. Along this large pressure range, both the
 condensate fraction  and the roton energy decrease but do
 not become zero. The roton energies obtained are compared with recent
 experimental data in the overpressurized regime. 
\end{abstract}


\maketitle

Quantum fluids in metastable states are presently a research topic of
fundamental interest from both the experimental and theoretical
viewpoints~\cite{budapest}. 
The extremely low temperature achieved in liquid helium makes
this liquid the most pure in nature and therefore the optimal choice for 
observing homogeneous nucleation, which is an intrinsic property of the
liquid. Caupin, Balibar and collaborators have studied profusely the
negative pressure regime by focusing high intensity ultrasound bursts in
bulk helium~\cite{caupin,balibar}. In liquid $^4$He they have measured a negative pressure of
-9.4 bar, only 0.2 bar above the spinodal point predicted by microscopic
theory~\cite{spinodal,pimc1}.
The same experimental team has used recently this acoustic technique to
pressurize small quantities of liquid $^4$He up to 160 bar at temperatures 
0.05 K $<T <1$ K~\cite{werner}. This
pressure is the highest pressure ever realized in overpressurized liquid
$^4$He and is much larger than the liquid-solid equilibrium pressure, which
at $T=0$ K is 25.3 bar. In the experimental setup used by Werner \textit{et
al.}~\cite{werner}, the overpressurized regime has become accessible by avoiding 
nucleation on the walls of the
container and therefore allowing only homogeneous nucleation. This
nucleation was not observed along this large increase of the liquid pressure beyond
the freezing point.

Liquid $^4$He in metastable states has also been obtained by immersing it
in different porous media. Albergamo \textit{et al.}~\cite{albergamo} have carried out
neutron scattering experiments in a medium with 47 \AA\ pore diameter filled
with densities below the equilibrium density and negative pressures up to
-5 bar. The achievement of negative pressures in this medium is
attributed to the stretching that the liquid supports due to the strong 
attraction to the walls. Using a different material, with 44 \AA\ pore
diameter, Pearce \textit{et al.}~\cite{glyde} have reported neutron scattering data in
the high density regime observing a liquid phase up to $\sim$40 bar.
Therefore, the confinement of helium in porous media makes feasible 
extensions of the pressure on both sides of the stable liquid phase.
The nature of these two metastable regions
presents a significant difference. At negative pressure, there exists an end
point (spinodal point) where the speed of sound becomes zero, and it is
thermodynamically forbidden to cross it maintaining a homogeneous liquid
phase. On the contrary, such a point does not exist on the overpressurized side.    
Nevertheless, Schneider and Enz~\cite{schneider} suggested that the pressurized phase has
also an end point corresponding to the pressure where the excitation energy
of the roton might vanish. On the other hand, and according to Jackson
\textit{et al.}~\cite{jackson} and Halinen \textit{et al.}~\cite{halinen}, the vanishing of the roton
energy is not the instability that causes the liquid-solid transition in
$^4$He but another one involving a 6-fold symmetric soft mode in the
two-body correlations.

The roton mode is not observed anymore when liquid $^4$He crystallizes.
In the overpressurized regime, between 25 and 38.5 bar, the neutron-scattering data of 
Ref.~\cite{glyde} show the roton excitation, while the maxon disappears, 
probably due to the fact that its
corresponding energy might exceed twice the roton energy. At the liquid-solid
transition, the roton energy is still finite.
The latter feature means that
$^4$He is superfluid when it crystallizes. This result seems to be in
disagreement with a recent experimental work by Yamamoto \textit{et
al.}~\cite{yamamoto}
who reported a critical temperature $T_{\text c}=0$ at a pressure $P \sim
35$ bar in a porous material. As a first explanation, the differences
observed in the two experiments can be attributed to the smaller pore diameter 
(25 \AA) used by Yamamoto \textit{et al.}~\cite{yamamoto}, but additional work would be
necessary to confirm this argument. It is worth noticing that the findings
of Ref.~\cite{yamamoto} imply the existence of a superfluid to normal phase
transition at zero temperature, an intriguing possibility that makes the
overpressurized regime even more interesting. In this direction,
Nozi\`eres~\cite{nozieres} has
predicted recently that the condensate fraction could vanish at a certain
pressure, and therefore a normal liquid before freezing would be possible.
     
At present, the theoretical knowledge of the metastable regime at negative 
pressures is rather complete with an overall agreement among quantum Monte
Carlo, hypernetted chain (HNC) based on Euler-Lagrange optimization, and density 
functional approaches~\cite{budapest}. On the contrary, the pressurized liquid remains until now
nearly unexplored. In the present work, we present 
diffusion Monte Carlo (DMC) results for the overpressurized phase up to 
$P\sim 275$ bar. The DMC method is probably the best suited way to deal with this 
metastable regime since the physical phase of the system is controlled by
the trial wave function used for importance sampling. Our results show a
superfluid phase in which, up to the highest pressure studied, the condensate 
fraction is small but discernible and the excitation energy of the roton decreases
but  does not reach the zero limit.

The DMC method is nowadays a well-known tool devised to study quantum fluids and
solids at zero temperature. Its starting point is the Schr\"odinger equation written
in imaginary time,
\begin{equation}
-\hbar \frac{\partial \Psi(\bm{R},t)}{\partial t} = (H- E_{\text r}) \Psi(\bm{R},t) \ ,
\label{srodin}
\end{equation}
with an $N$-particle Hamiltonian
\begin{equation}
 H = -\frac{\hbar^2}{2m} \sum_{i=1}^{N} \bm{\nabla}_i^2 + \sum_{i<j}^{N} V(r_{ij})  \ .
\label{hamilto}
\end{equation}
In Eq. (\ref{srodin}), $E_{\text r}$ is a constant acting as a reference energy and
$\bm{R} \equiv (\bm{r}_1,\ldots,\bm{r}_N)$ is a \textit{walker} in Monte
Carlo terminology. DMC solves stochastically the Schr\"odinger
equation (\ref{srodin}) replacing $\Psi(\bm{R},t)$ by $\Phi(\bm{R},t)=
\Psi(\bm{R},t) \psi(\bm{R})$
with $\psi(\bm{R})$ a trial wave function used for importance sampling.
When $t \rightarrow \infty$ only the lowest energy eigenfunction, not
orthogonal to $\psi(\bm{R})$, survives. The simulation of the liquid in its
ground state is carried out by using a Jastrow approximation,
$\psi_{\text J}(\bm{R}) = \prod_{i<j}^{N} f(r_{ij})$.
As in our previous work~\cite{boro1} on the equation of state of liquid $^4$He in the
stable regime, $f(r)$  is a model proposed by
Reatto~\cite{reatto} which incorporates nearly optimal short and medium range two-body
correlations. As a matter of comparison, simulations of the crystalline
hcp phase have been also carried out; in this case, a Nosanow-Jastrow model
is used, $\psi_{\text{NJ}}(\bm{R}) = \psi_{\text J}(\bm{R}) 
\prod_{i}^{N}g(r_{iI})$, with $g(r)$ a gaussian function linking every
particle $i$ to a fixed point $\bm{r}_I$ of the lattice. 
  
The calculation of the roton energy using the DMC method is more involved
since it corresponds to an excited state. In this case, we have taken as
a trial wave function for importance sampling an eigenstate of the total
momentum operator which incorporates backflow correlations, as originally
proposed by Feynman and Cohen~\cite{feynman},
\begin{equation}
 \psi_{\text{BF}}(\bm{R}) = \sum_{i=1}^{N} e^{i \bm{q} \cdot \tilde{\bm{r}}_i} \,
 \psi_{\text J}(\bm{R}) \ ,
 \label{feycohen}
 \end{equation} 
with $\tilde{\bm{r}}_i = \bm{r}_i + \sum_{j \ne i}^{N} \eta(r_{ij})
\bm{r}_{ij}$, and $\eta(r)=\lambda \exp [ -((r-r_{\text b})/\omega_{\text
b})^2 ]$. The only objective of our calculation is the energy of the
collective mode, which is the same for excitations with momenta $\bm{q}$ and
$-\bm{q}$. Therefore, we avoid the complexity of working with a complex wave
function (\ref{feycohen}) by using a superposition of both states.
Proceeding in this form, the excited state  turns into a fermion-like
problem since the trial wave function is real but not positive
everywhere~\cite{boro2}.
In a first step, we have used the fixed-node (FN) approximation, which provides an
upper bound to the roton energy. We verified that the introduction of
backflow correlations in the trial wave function produces results
quite close to experimental data at the equilibrium
density, especially near the roton minimum. The nodal constraint imposed by
FN is removed, in a second step, by using the released-node (RN) technique
in which the walkers are allowed to cross the nodal surface imposed by
$\psi$ and survive for a finite lifetime.  

\begin{figure}
\centering
        \includegraphics[width=0.8\linewidth]{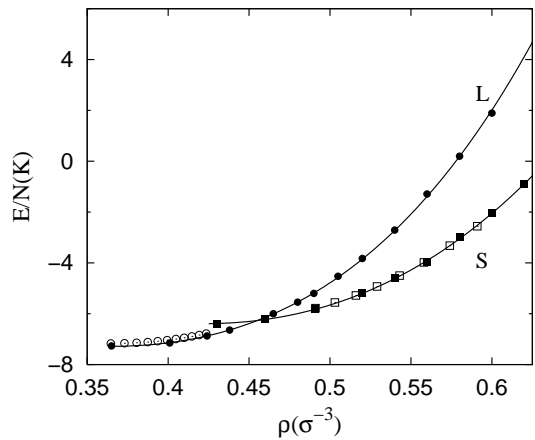}%
        \caption{Energy per particle of liquid $^4$He from the equilibrium 
	density up to the highest density calculated, 0.6 $\sigma^{-3}$,
	(solid circles). The solid line corresponds to the fit to the DMC
	energies using Eq. (\protect\ref{eqestat}), and the open circles are experimental data
	in the stable regime from Ref.~\protect\cite{exp1}. DMC results for the solid phase
	are shown as solid squares and compared with experimental data from
	Ref.~\protect\cite{exp2} (open squares). The error bars of our data
	are smaller than the size of the symbols.  }
\end{figure}

In Fig. 1, we report the results obtained for the energy per particle of
the liquid phase as a function of the density. The He-He interaction
corresponds to the  HFD-B(HE) Aziz potential~\cite{aziz} and the number of atoms used
in the simulation increases with density: $N=150$ near freezing and $N=250$
at the highest densities; periodic boundary conditions are assumed in all the 
simulations. Possible biases coming from the use of a finite
number of walkers and from a finite time-step in the employed second-order algorithm 
are reduced to the level of the statistical noise. The DMC energies are
accurately parameterized, from the spinodal point up to the highest
densities calculated, by the analytical form
\begin{eqnarray}
e(\rho) & =  & e_0 +e_1 \left( \rho/\rho_c -1 \right) \left( 1 - \left(
\rho/\rho_c -1 \right) \right)  \nonumber \\
  &  & + b_3 (\rho/\rho_c- 1)^3 + b_4(\rho/\rho_c -1)^4 \ ,\label{eqestat}
\end{eqnarray}
with $e=E/N$, and $\rho_c=0.264\ \sigma^{-3}$ ($\sigma=2.556$ \AA) the
spinodal density. The rest of parameters in Eq. (\ref{eqestat}) are 
$e_0=-6.3884(40)$ K, $e_1=-4.274(31)$ K, $b_3=1.532(12)$ K, and
$b_4=1.433(24)$ K, the figures in parenthesis being the
statistical errors~\cite{qfs}. DMC results of the energies of the solid phase,
calculated using the Nosanow-Jastrow trial wave function and an hcp
lattice, are also shown in Fig. 1. The comparison between the liquid and
solid phase simulations shows clearly that DMC is effectively able to study
the overpressurized liquid phase in spite of not being the ground-state
(minimum energy) configuration, which manifestly corresponds to the solid
phase beyond the freezing point. 

\begin{figure}
\centering
        \includegraphics[width=0.8\linewidth]{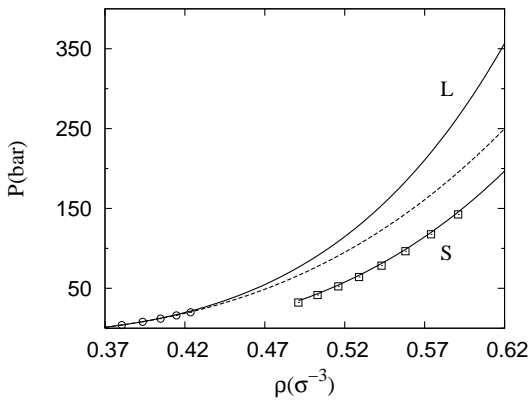}%
        \caption{Pressure as a function of the density. The solid lines stand
	for the DMC results obtained from the equations of state of the
	liquid and solid phases shown in
	Fig. 1. The dashed line is the extrapolation from experimental
	data~\protect\cite{werner,exp3}; the symbols correspond to experimental data for the
	liquid~\protect\cite{exp1} and
	solid phases~\protect\cite{exp2}. }
\end{figure}

Using the equation of state (\ref{eqestat}), the pressure is obtained from
its thermodynamic definition $P(\rho)=\rho^2(\partial e/ \partial \rho)$.
The results obtained,  which are shown in Fig. 2, reproduce accurately the 
experimental data~\cite{exp1} in the stable regime and predict a pressure $P \simeq
275$ bar at the highest density evaluated, $\rho=0.6\ \sigma^{-3}$. Our
results are compared in the same figure with the analytic form suggested 
in Ref.~\cite{werner}, adjusted to Abraham's experimental data~\cite{exp3}.
Below the freezing point, both curves agree but they give significantly different
values at higher densities; the difference amounts to $\sim 100$ bar at
$\rho=0.6 \ \sigma^{-3}$. As a matter of comparison, the figure also shows
the pressure of the solid phase, derived from the DMC equation of state
shown in Fig. 1. 

\begin{figure}
\centering
        \includegraphics[width=0.8\linewidth]{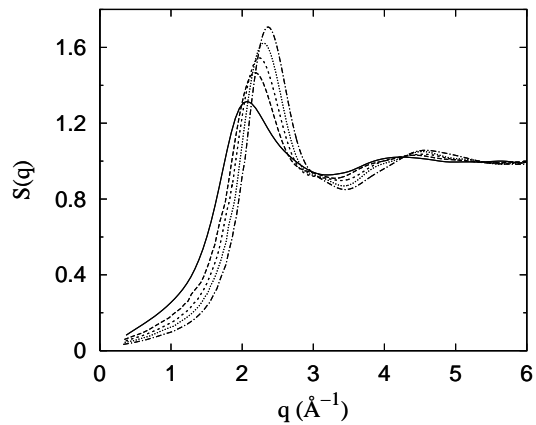}%
        \caption{Static structure function of the liquid phase for
	different densities. From bottom to top in the height of the main
	peak, the results correspond to densities 0.365, 0.438, 0.490,
	0.540, and 0.6 $\sigma^{-3}$. }
\end{figure}

A characteristic feature of a solid phase is the presence of
high-intensity peaks in the static structure function $S(q)=\langle
\rho_{\bm q} \rho_{-\bm q} \rangle / N$, $\rho_{\bm q}=\sum_{i=1}^{N} e^{i
{\bm q}\cdot {\bm r}_i} $, in the reciprocal lattice sites. Following the
overpressurized liquid phase, we have not observed this feature and thus 
the liquid nature of the system is confirmed. In Fig. 3, results of $S(q)$ for
densities ranging from equilibrium up to the highest densities studied are
reported. The results show the expected behavior: when $\rho$ increases, 
the strength of the main peak increases and moves to higher momenta  
in a monotonic way. At low momenta, the slope of $S(q)$ decreases with the
density, following the limiting behavior $\lim_{q \rightarrow 0} S(q)= \hbar
q/(2mc)$ driven by the speed of sound $c$.     

\begin{figure}
\centering
        \includegraphics[width=0.8\linewidth]{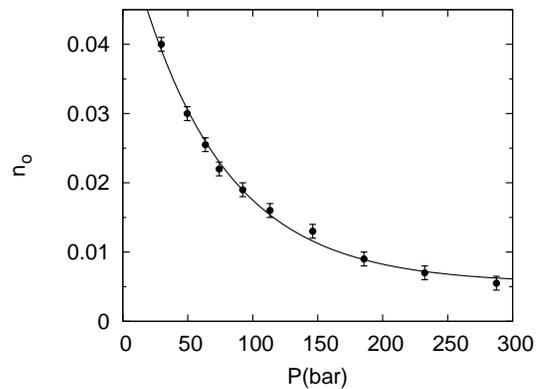}%
        \caption{Condensate fraction of liquid $^4$He in the
	overpressurized regime. The line is an exponential fit to the
	DMC results. }
\end{figure}

A characteristic signature of bulk superfluid $^4$He is a finite value of
its condensate fraction, i.e., the fraction of particles occupying the
zero-momentum state. As usual in a homogeneous system, the condensate
fraction $n_0$ has been estimated from the long range behavior of the
one-body density matrix, $\lim_{r \rightarrow \infty} \rho(r)=n_0$. 
The results obtained
for $n_0$, from the melting pressure up to nearly 300 bar, are plotted in
Fig. 4. The line on top of the data corresponds to an exponential fit which
reproduces quite accurately our DMC results ($n_0(P)\simeq A e^{-\beta P}$,
with $A=0.052$ and $\beta=0.015$ bar$^{-1}$). As one can see in the figure,
$n_0$ decreases quite fast until $P= 100$ bar and then the slope
decreases, approaching a value $n_0 \simeq 0.005$ at the highest density. With
the same procedure, we obtained~\cite{boro1} $n_0=0.084(1)$ at the equilibrium density $\rho_0=0.365\
\sigma^{-3}$, value which is compatible with PIMC estimations at low
temperature~\cite{pimcnk} (0.069(10) at $T=1.18$ K and 0.087(10) at $T=1.54$ K).    

\begin{figure}
\centering
        \includegraphics[width=0.8\linewidth]{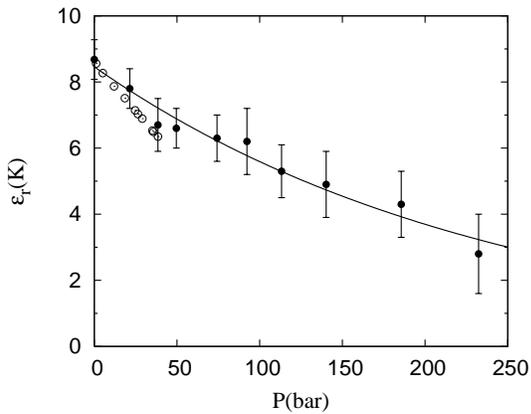}%
        \caption{Roton energy as a function of the density (solid circles).
	Open circles stand for experimental data from
	Ref.~\protect\cite{glyde}. The line is
	an exponential fit to the DMC data. }
\end{figure}

Superfluidity in bulk $^4$He is manifested in the dynamic structure
function by the clear signature of the roton collective mode. By increasing
the density the roton energy $\epsilon_{\text r}$ decreases, and its
approach to zero has been proposed as a possible final point in the
overpressurized regime~\cite{schneider}. Trying to discern this hypothesis, we have carried
out a RN calculation of the roton energy beyond the freezing point. The same
methodology was used in the past in a DMC calculation of the phonon-roton 
spectrum at equilibrium and freezing densities~\cite{boro2} arriving at an accurate
description of the experimental data. The results obtained are shown in
Fig. 5. Experimental data obtained by neutron scattering experiments on
superfluid $^4$He in a porous media and up to 40 bar are also plotted in
the figure~\cite{glyde}. From 0 to 40 bar, $\epsilon_{\text r}$ decreases linearly with
the pressure and our data reproduces well this behavior. However,
increasing the pressure this slope is reduced and, at the highest density
studied the roton energy is still different from zero ($\epsilon_{\text
r}=2.8 \pm 1.2$ K at $\rho=0.58\ \sigma^{-3}$). It is worth noticing that the statistical errors are rather large
and difficult to reduce since $\epsilon_{\text r}= E_{\text r} - E_0$, with
$E_{\text r}$ and $E_0$ the total energy of the excited and ground state,
respectively. The excited state energy is estimated through an exponential
fit $E(t)=E_{\text r} + A e^{-(t/\tau)}$, with $t$ the released time. The
uncertainty of this extrapolation is under control since in all cases the
difference between considering the last calculated point in released time
or $E_{\text r}$ is of the same order as the statistical noise. At each
density, the number of particles has been adjusted to be as close to the
roton momentum as possible; nevertheless, as only discrete values of ${\bm
q}$ are accessible, possible corrections to the energy due to this fact 
are estimated to be less than 0.5 K in all cases. The obtained values of 
the roton momentum do not follow linear dependence with pressure, 
as the data in the stable liquid regime might suggest~\cite{gibbs}. 
As the pressure is increased, the slope of the roton momentum as a function 
of pressure decreases.  

In summary, the present DMC study of superfluid $^4$He in the
overpressurized regime does not show any signature of final (spinodal) point.
The structure factor, condensate fraction, and roton energy are mainly driven by
the density and not by the pressure. As shown in Figs. 1 and 2, 
equal changes in density in the stable and metastable regimes produce
different changes in pressure.
This leads, in the density range
studied, to an approximated exponential behavior with pressure of
magnitudes like $n_0$ and $\epsilon_{\text r}$.

We thank M. Barranco, F. Caupin, J. Navarro, and M. Saarela for useful
discussions. Partial financial support from DGI (Spain) Grant No. BFM2002-00466 
and Generalitat de Catalunya Grant No. 2001SGR-00222 is gratefully
acknowledged. We acknowledge the support of Central Computing Services at
the Johannes Kepler University in Linz, where part of the computations were
performed.

\end{document}